\documentclass[aps,prl,twocolumn,longbibliography]{revtex4-1}
\usepackage{epsfig,amsopn}
\usepackage{graphicx}
\usepackage{color}
\usepackage{amsmath,amssymb}
\usepackage{array}
\usepackage{enumerate}
\newcommand\bea{\begin{eqnarray}}
\newcommand\eea{\end{eqnarray}}
\newcommand\beq{\begin{equation}}
\newcommand\eeq{\end{equation}}

\def\nn{\nonumber}
\def\f{\frac}
\def\al{\alpha}

\def\ga{\gamma}

\def\si{\sigma}

\def\De{\Delta}

\def\ua{\uparrow}
\def\da{\downarrow}

\def\th{\theta}

\begin{document}
\title{Josephson diode effect in one-dimensional quantum wires connected to superconductors with mixed singlet-triplet pairing}
\author{ Abhiram Soori}  
\email{abhirams@uohyd.ac.in}
\affiliation{ School of Physics, University of Hyderabad, Prof. C. R. Rao Road, Gachibowli, Hyderabad-500046, India}

\begin{abstract}
The Josephson diode effect (JDE), characterized by asymmetric critical currents in a Josephson junction, has drawn considerable attention in the field of condensed matter physics. We investigate the conditions under which JDE can manifest in a one-dimensional Josephson junction composed of a spin-orbit-coupled quantum wire with an applied Zeeman field, connected between two superconductors. Our study reveals that while spin-orbit coupling (SOC) and a Zeeman field in the quantum wire are not sufficient to induce JDE when the superconductors are purely singlet,  introduction of triplet pairing in the superconductors leads to the emergence of JDE. This finding highlights the potential of JDE as a probe for triplet superconductivity. We further demonstrate that even in absence of SOC in the quantum wire, JDE can arise when the directions of the triplet pairing and the Zeeman field are non-collinear, provided the superconductors exhibit mixed singlet-triplet pairing. Additionally, we identify specific conditions under which JDE is absent, namely, when the pairing is purely triplet and the directions of the SOC and the triplet pairing are perpendicular. Our findings indicate that JDE is always accompanied by  anomalous Josephson effect. The diode effect coefficient is found to oscillate with variations in the chemical potential of the quantum wire, driven by Fabry-P\'erot interference effects. Our results suggest that quantum wires  connected across superconductors can serve as effective platforms for probing triplet superconductivity through the observation of JDE.
\end{abstract}
\maketitle

{\it Introduction .-} When the phases of two superconductors (SCs) separated by a thin insulator differ, a current flows from one to the other, a phenomenon known as Josephson effect named after its discoverer~\cite{josephson62,furusaki99}. The dependence of such current on the difference in the phases is called current phase relation (CPR).  Josephson diode effect (JDE), a phenomenon in Josephson junctions  is characterised by unequal magnitudes of the maximum and minimum values of currents in CPR. JDE has attracted  attention of theorists and experimentalists in the recent years~\cite{wu2022,baum22mca,yuan2022,costa23,soori23aje,soori23njde,soori24bam,dutta24,fu2024diode,fracassi2024,yerin2024}. SOC  along with Zeeman field in a metallic region connected to superconductors on either sides is known to show JDE due to magnetochiral anisotropy in two-dimensional systems~\cite{baum22mca,Turini2022,costa23,rein2024}. We explore - ``{\it whether  JDE is possible in purely one-dimensional Josephson junction comprising of a spin-orbit coupled quantum wire with an applied Zeeman field at the center?}' 

When a Zeeman field is applied to a spin-orbit-coupled quantum wire, parallel to the direction of the SOC, it results in different velocities for the left-mover and right-mover at any given energy, a phenomenon known as magnetochiral anisotropy.  However, merely connecting such a quantum wire to s-wave superconductors on either side is not sufficient to induce the JDE. To observe JDE in one-dimensional systems, the necessary ingredients include: SOC in the two superconductors, and Zeeman field components both parallel and transverse to the SOC in the quantum wire.  Another way to induce JDE is by introducing a width to the spin-orbit-coupled quantum wire.  A paper by Meyer and Houzet discusses some of these points in detail~\cite{meyer24diode}.   Additionally, Majorana fermions, which appear in spin-orbit-coupled quantum wires, are known to enhance JDE~\cite{cayao24diode}. Against this backdrop, we investigate CPR of Josephson junctions between superconductors with mixed singlet-triplet (or purely triplet) pairing, connected by a single-channel one-dimensional spin-orbit coupled quantum wire on which a Zeeman field is applied. 

Josephson junctions involving triplet superconductors with a ferromagnet in the middle are known to exhibit the anomalous Josephson effect~\cite{liu2014}. This phenomenon occurs when the $\vec{d}$-vectors of the two superconductors and the spin polarization direction of the ferromagnet are non-coplanar. Noncentrosymmetric superconductors, such as CePt$_3$Si and $\alpha$-BiPd, are known to host both singlet and triplet pairings simultaneously in the same material~\cite{klam2014,xu2020}. SOC is also known to facilitate long-range triplet superconductivity~\cite{jacobsen2015}, providing us with numerous materials in which triplet pairing can exist. We show that the triplet pairing in SCs can result in JDE along with  anomalous Josephson effect, which would have been absent if the pairing was purely singlet. 

{\it Calculations .-}
The system under study is a superconductor-quantum wire-superconductor junction, where the superconductors exhibit both singlet and triplet pairings. The central quantum wire features SOC, and a Zeeman field is applied parallel to the SOC. The Hamiltonian describing this system is given by
\bea 
H &=& -t\sum_{n=1}^{L_{sqs}-1}[\Psi^{\dagger}_{n+1}\tau_z\Psi_n+{\rm h.c.}] \nn \\ && -\mu_s\Big[\sum_{n=1}^{L_s} + \sum_{n=L_{sq}+1}^{L_{sqs}}\Big]\Psi^{\dagger}_n\tau_z\Psi_n -\mu_0\sum_{L_s+1}^{L_{sq}}\Psi^{\dagger}_n\tau_z\Psi_n \nn \\ &&  -\De_s\sum_{n=1}^{L_s}\Psi^{\dagger}_n(\cos{\phi} \tau_y\si_y+\sin{\phi} \tau_x\si_y)\Psi_n \nn \\&&- \f{i\De_t}{2} \sum_{n=1}^{L_s-1}[\Psi^{\dagger}_{n+1}(\cos{\phi}\tau_x\si_{\th}+\sin{\phi}\tau_y\si_{\th})\Psi_{n} -{\rm h.c.}] \nn \\&& + b \sum_{n=L_{s}+1}^{L_{sq}} \Psi^{\dagger}_n\tau_z\si_z\Psi_n -\f{i \al}{2}\sum_{n=L_{s}+1}^{L_{sq}-1}  [\Psi^{\dagger}_{n+1}\si_z\Psi_n-{\rm h.c.}] \nn \\ &&-\De_s\sum_{n=L_{sq}+1}^{L_{sqs}}\Psi^{\dagger}_n  \tau_y\si_y \Psi_n \nn \\ && - \f{i\De_t}{2} \sum_{n=L_{sq}+1}^{L_{sqs}-1}[\Psi^{\dagger}_{n+1}\tau_x\si_{\th}\Psi_{n}-{\rm h.c.}],  
\eea
where $L_s$ is the number of sites in each  superconductor, $L_q$ is the number of sites in the central quantum wire, $L_{sq}=L_s+L_q$, $L_{sqs}=2L_s+L_q$, $\Psi_n=[c_{n,\ua},~c_{n,\da},~c^{\dagger}_{n,\ua},~c^{\dagger}_{n,\da}]^T$, $c_{n,\si}$ annihilates an electron of spin-$\si$ at site $n$,  $\tau_{x,y,z}$ are Pauli spin matrices that act on the particle-hole space, $\si_{x,y,z}$ are Pauli spin matrices that act on the spin space, $\si_{\th}=\si_z\cos{\th}-\si_x\sin{\th}$, $\th$ is the angle between the direction of SOC in the central quantum wire and the direction of triplet pairing in the superconductor, $\phi$ is the difference between the phases of the superconducting pair potentials on the two SCs, $t$ is the hopping amplitude taken to be same in the entire system, $\mu_s$ ($\mu_0$) is the chemical potential on the SC (quantum wire), $\De_s$ ($\De_t$) is the magnitude of singlet (triplet) pairing amplitude in the two superconductors, $b$ is the energy scale associated with the Zeeman energy and $\al$ is the strength of SOC. The Hamiltonian can be expressed as a matrix of size $4L_{sqs}\times 4L_{sqs}$ and numerically diagonalised. We start with the numerical diagonalization of the Hamiltonian as $\phi\to 0^{+}$ . The negative energy states are considered fully occupied, with the positive energy states left unoccupied. As $\phi$ increases incrementally, the occupied states are those states that evolve incrementally from the filled negative energy states at $\phi=0^+$.

The charge is conserved in the quantum wire, and hence the charge current can be calculated in the quantum wire. We calculate the charge current using current operator defined at the bond connecting the quantum wire to the superconductor defined by $\hat J = it e(\Psi^{\dagger}_{L_{s}+1}\Psi_{L_s}-{h.c.})$. Current carried by the occupied states are summed over to get the total current $J$. 
 
\begin{figure}[htb]
\includegraphics[width=8cm]{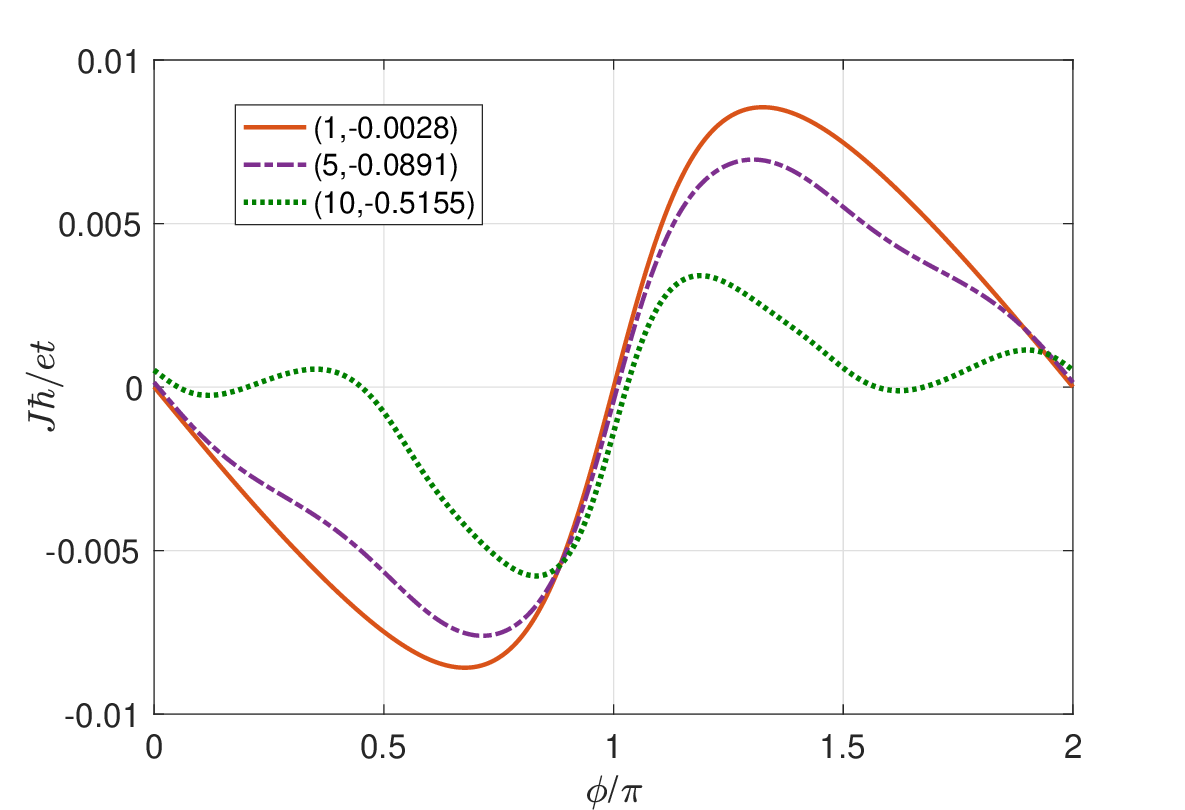}
\caption{CPR. Legend shows the values of $(\De_t/\De_s,\ga)$ for each curve. Parameters: $\mu_s=\mu_0=-1.875t$, $\al=0.05t$, $\De_s=0.0125t$, $b=0.015t$, $\th=0$, $L_s=L_q=20$.}\label{fig:cpr}
\end{figure}

{\it Results .-}
We performed the calculations using parameter values close to those found in experiments~\cite{rainis2013,Lutchyn2018,cayao24diode}. Parameters take following realistic values for InAs or InSb quantum wires~\cite{rainis2013,Lutchyn2018,cayao24diode}: $t=40meV$ (which can be calculated from $t=\hbar^2/2m^*a^2$ and using the values $m^*=0.02m_e$ for the effective mass, $m_e$ being the electron mass and $a=10nm$, the effective lattice spacing), $\mu_s=\mu_0=-1.875t$ (but $\mu_0$ can be varied by application of a gate voltage to the central quantum wire), $\al=2meV=0.05t$, $\De_s=0.5meV=0.0125t$ (for superconductors like NbTiN), $\De_t=0.5meV=0.0125t$ (for BiPd~\cite{Sun2015}, but can vary depending on the choice of superconductor), Zeeman energy has a magnitude of $b=0.6meV=0.015t$ for a magnetic field of $1T$ (assuming a value of 10 for the $g$-factor). 
Fig.~\ref{fig:cpr} shows CPR for different values of the triplet pairing amplitude, $\Delta_t$, while keeping $\mu_s=\mu_0=-1.875t$, $\alpha=0.05t$, $\Delta_s=0.0125t$, $b=0.015t$, $\theta=0$, and $L_s=L_q=20$ fixed. The figure legend displays the corresponding values of $(\Delta_t/\Delta_s, \gamma)$. As $\Delta_t$ increases, the diode effect coefficient $\gamma$ also increases in magnitude. In the limit $\Delta_t \to 0$, $\gamma$ approaches zero, indicating that the diode effect is rooted in a nonzero triplet pairing amplitude, $\Delta_t$. JDE is always accompanied by anomalous Josephson effect in our study. 

\begin{figure}[htb]
\includegraphics[width=4cm]{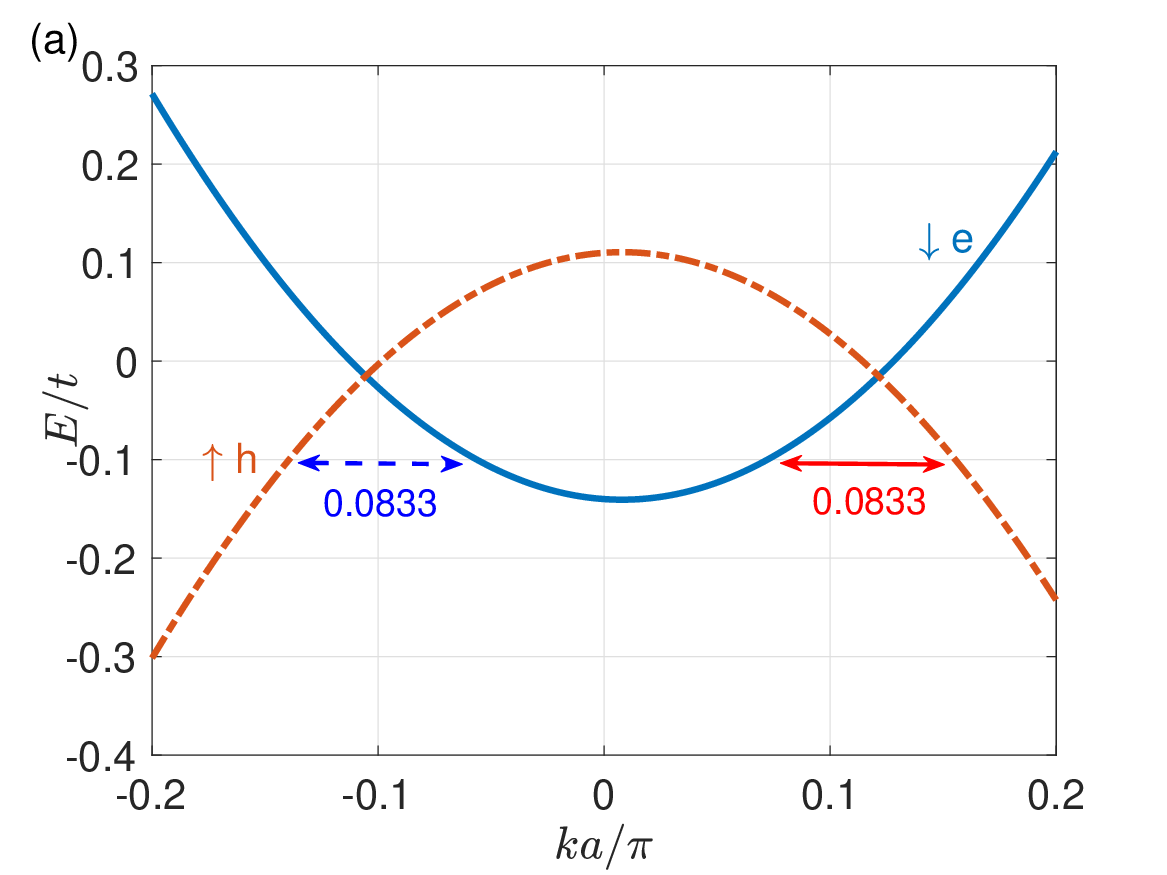}
\includegraphics[width=4cm]{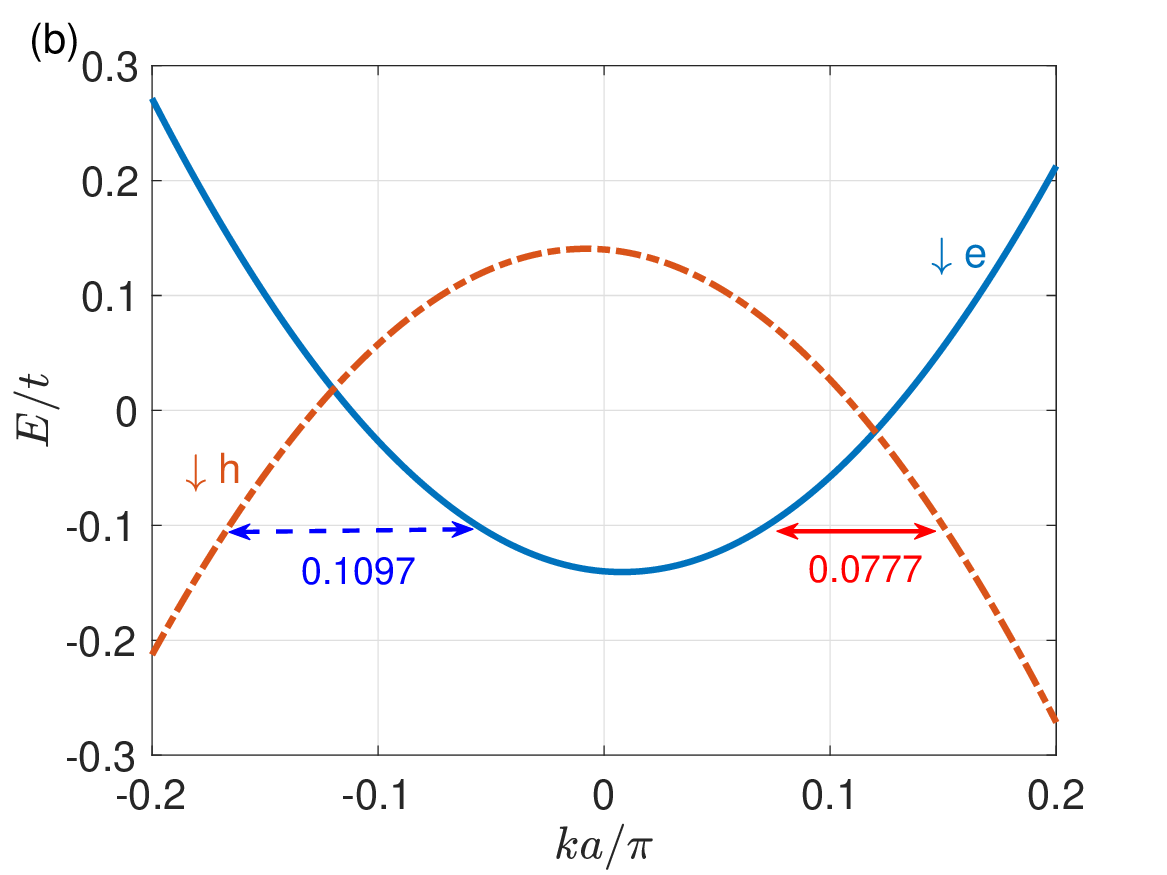}
\caption{ Dispersion of the central quantum wire in the sectors (a)$\da$-spin electron, $\ua$-spin hole  (b)$\da$-spin electron, $\da$-spin hole. It can be seen that the dynamical phases picked up by the processes that carry current in forward and backward directions are same [different]  in (a) [(b)].  Parameters: $\mu_0=-1.875t$, $\al=0.05t$, $b=0.015t$. }\label{fig:phase-eh}
\end{figure}

In the pure singlet phase of the superconductor, the diode effect is absent. This is because the modes that carry Josephson current in the central quantum wire can be decomposed into two sectors: (i) up-spin electron and down-spin hole, and (ii) down-spin electron and up-spin hole. In each sector, the dynamical phase accumulated by the pair of states—right-moving electron and left-moving hole (which carry current in the forward direction)—is the same as that accumulated by the left-moving electron and right-moving hole (which carry current in the backward direction), as shown in Fig.~\ref{fig:phase-eh}(a). These two pairs of states carry currents in opposite directions, resulting in the absence of the diode effect when the superconductors are purely in the singlet phase.

However, when $\Delta_t \neq 0$ (with $\theta=0$, indicating that the pairing is $(|\uparrow\uparrow\rangle - |\downarrow\downarrow\rangle)$), triplet pairing between electrons and holes of the same spin becomes possible. This allows the electron-hole pairs of states with the same spin to carry the Josephson current. Unlike in the singlet case, the pairs of states—left-moving electron and right-moving hole, and right-moving electron and left-moving hole, all with the same spin—do not accumulate the same dynamical phase during one   back-and-forth journey, as illustrated in Fig.~\ref{fig:phase-eh}(b). This  difference in phases accumulated  leads to JDE when the triplet pairing amplitude is nonzero.

The CPR can be expressed as a sum of different harmonics by 
\beq 
J(\phi) = \sum_{n=1}^{\infty}[J_{s,n}\sin{n\phi}+J_{c,n}\cos{n\phi}], \label{eq:harmo}
\eeq
where $n$ takes positive integer values. Typically, the first few terms dominate CPR, and as $n$ becomes larger, $J_{s,n}$ and $J_{c,n}$ become negligibly small. For the CPR in Fig.~\ref{fig:cpr}, we list out the amplitudes of different harmonics in the table below. For CPR to exhibit anomalous Josephson effect,  $J_{s,m}$ and $J_{c,m}$ for a particular $m$ should be nonzero. For JDE to occur, any three out of $J_{s,2l}$,  $J_{c,2l}$, $J_{s,2m+1}$ and $J_{c,2m+1}$ (for some positive integer $l$ and nonnegative integer $m$) should be nonzero. 
It can be seen from Table~\ref{tab} that the $J_{s,1}$, $J_{c,1}$ and $J_{s,2}$ are nonzero and significant, which enables the diode effect to emerge.
\begin{table}[htb]
\begin{center}
  \begin{tabular}{ | l | c | r |}
    \hline
    $~n~$ & $J_{s,n}$ & $J_{c,n}$ \\ \hline
    ~1~ & $-0.0019$ &  $0.0019$ \\ \hline
    ~2~ &  $0.0022$ & $-1.49\times 10^{-4}$  \\ \hline
    ~3~& $-0.0015$ & $-4.72\times 10^{-5}$   \\ \hline 
    ~4~& $1.54\times 10^{-4}$ & $4.7\times 10^{-5}$   \\
    \hline
  \end{tabular}
\end{center}
\caption{Different harmonics in the CPR for $\De_t=10\De$ in Fig~\ref{fig:cpr}. See Eq.~\eqref{eq:harmo} for the meaning of $J_{s,n}$ and $J_{c,n}$. }\label{tab}
\end{table}

\begin{figure}[htb]
\includegraphics[width=4cm]{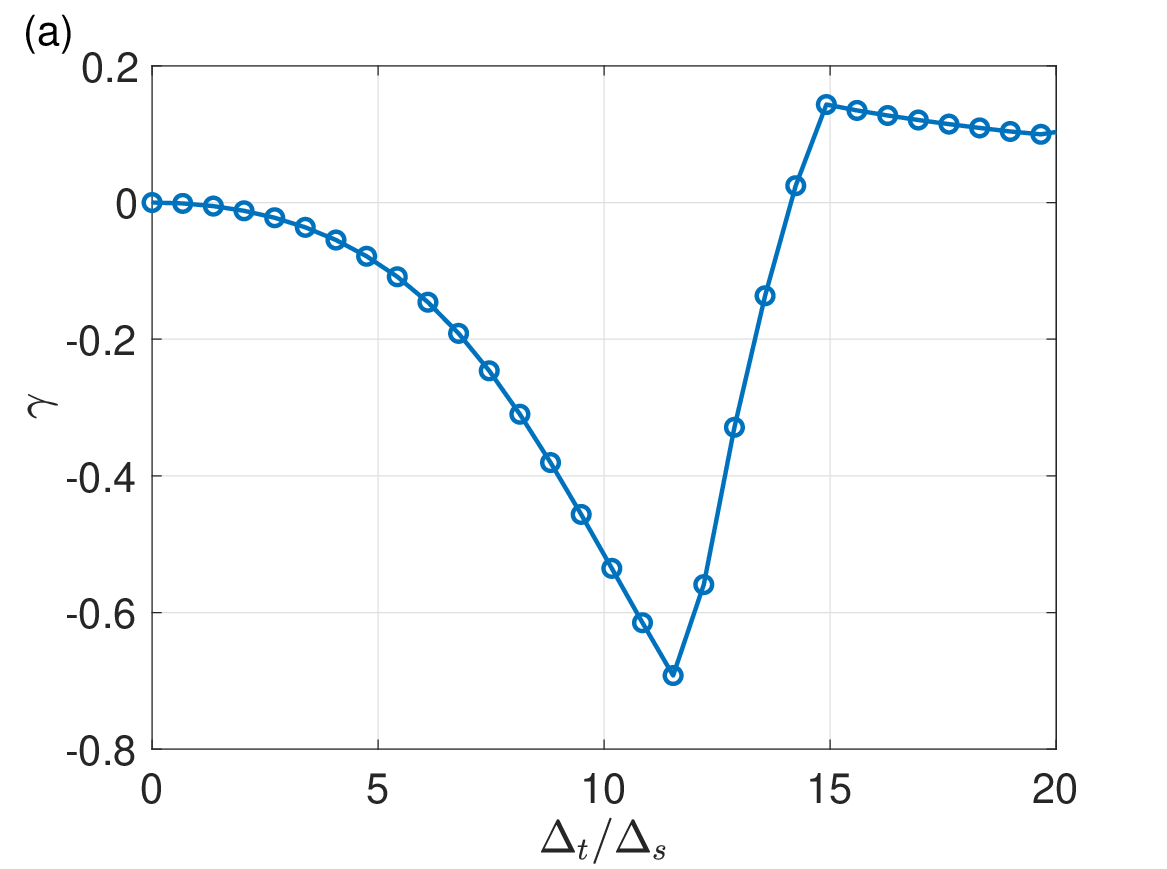}
\includegraphics[width=4cm]{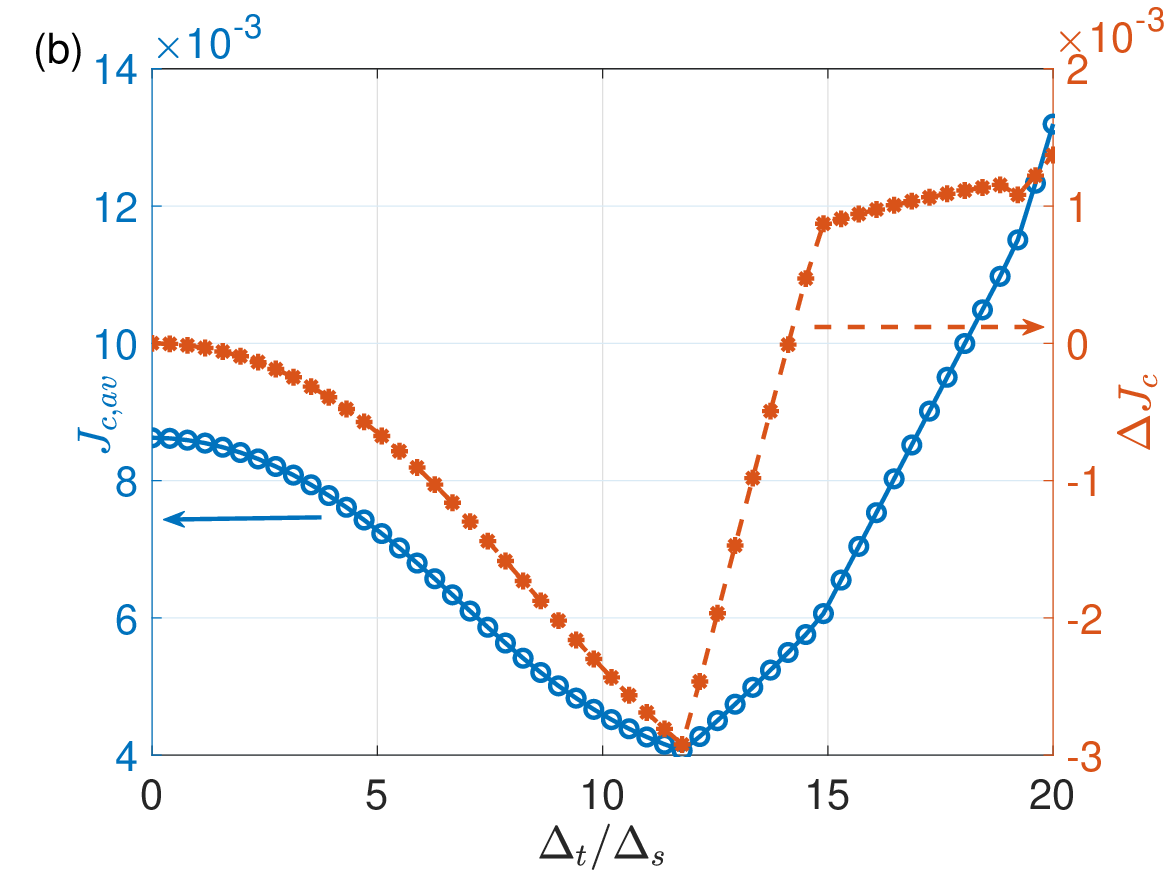}
\caption{(a) Diode effect coefficient $\ga$ versus $\De_t/\De_s$ for $\De_s=0.0125t$. (b) Average critical current $J_{c,av}$ and   the difference $\De J_c$ versus $\De_t/\De_s$. Parameters: $\mu_s=\mu_0=-1.875t$, $\al=0.05t$, $b=0.015t$, $\th=0$, $L_s=L_q=20$.}\label{fig:ga-Det}
\end{figure}

To explore how the diode effect coefficient $\gamma$ depends on the triplet pairing amplitude, we plot $\gamma$ against $\Delta_t/\Delta_s$ in Fig.~\ref{fig:ga-Det}(a) using the same parameter set. The figure shows that $\gamma$ initially grows in magnitude as $\Delta_t$ increases, reaching a peak before it sharply decreases. Since the diode effect arises from a nonzero triplet pairing amplitude, $|\gamma|$ first increases with $\Delta_t$. For small triplet pairing amplitudes, the Josephson current in CPR can be approximated as the sum of currents from singlet and triplet pairings. The CPR for singlet pairing alone takes the form $J = -J_{cs}\sin{\phi}$, while for triplet pairing alone, it is $J = J_{ct}\sin{(\phi+\phi_0)}$, where $\phi_0 \ll \pi$ and $J_{ct}, J_{cs} > 0$. This results in the combined CPR being $J \approx J_{ct}\sin{\phi} - J_{cs}\sin{\phi}$. Consequently, the average critical current $J_{c,av}$ decreases as the triplet pairing amplitude grows. But the difference between the critical currents $\Delta J_c$ in the forward and backward directions increases in magnitude with $\De_t$. This leads to a sharp rise in the diode effect coefficient $\gamma = \Delta J_c / J_{c,av}$ as $\Delta_t$ increases. In Fig.~\ref{fig:ga-Det}(b), we illustrate both the average critical current $J_{c,av}$ and the difference $\Delta J_c$ between the critical currents in  two directions. Once $\Delta_t$ surpasses a critical threshold, the triplet pairing becomes dominant, increasing the Josephson critical current. As $\Delta_t$ continues to rise, the superconducting gap increases and a competing mechanism begins to offset this increase. More states from the quantum wire contribute to the supercurrent as they lie within the gap. These states, existing at various energies, introduce varying contributions to the diode effect. While certain energy states favor a stronger Josephson current in the forward direction, this preferred direction reverses as additional states become involved. This causes a reduction in  magnitude of the difference $\Delta J_c$, leading to a decrease in the diode effect coefficient's magnitude. This explains the observed peak in magnitude of the diode effect coefficient as a function of $\Delta_t$.

\begin{figure}[htb]
\includegraphics[width=4cm]{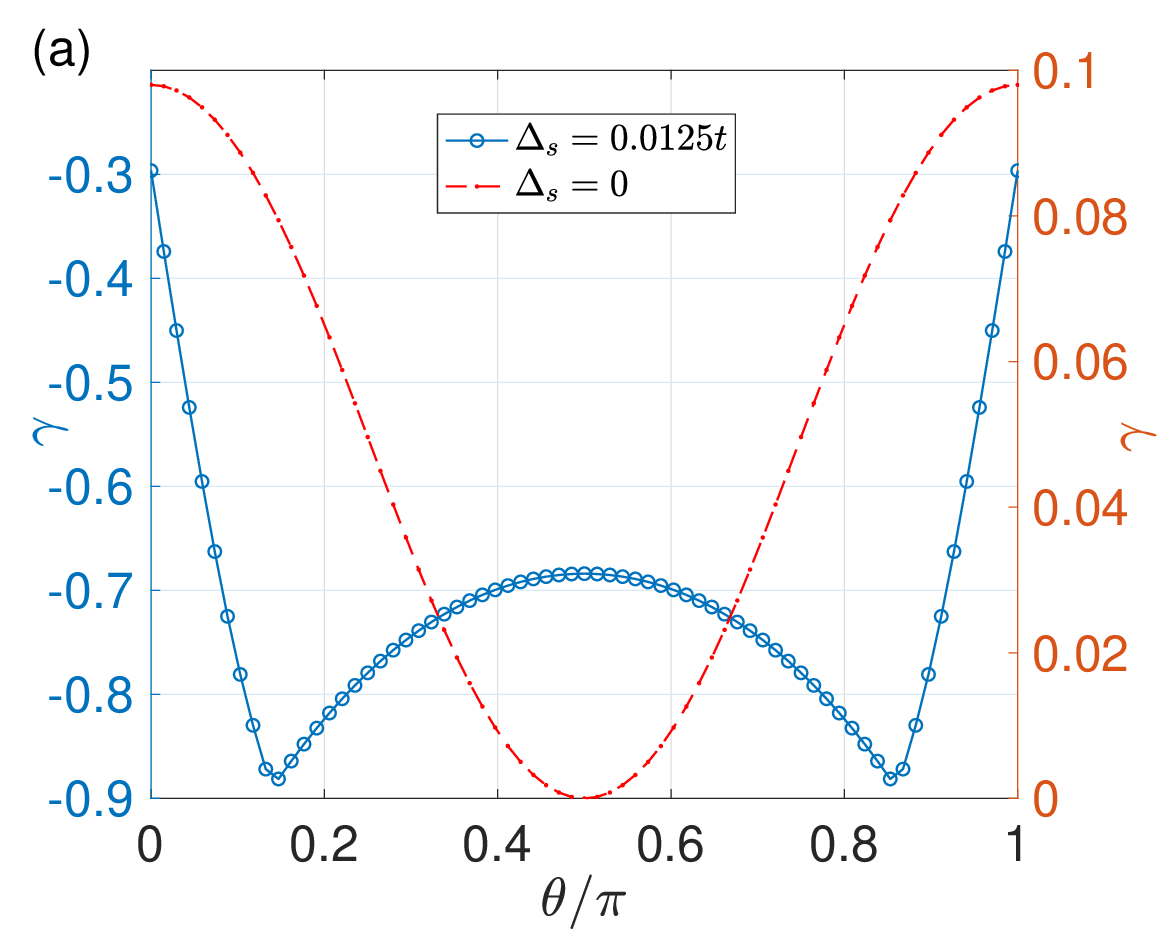}
\includegraphics[width=4cm]{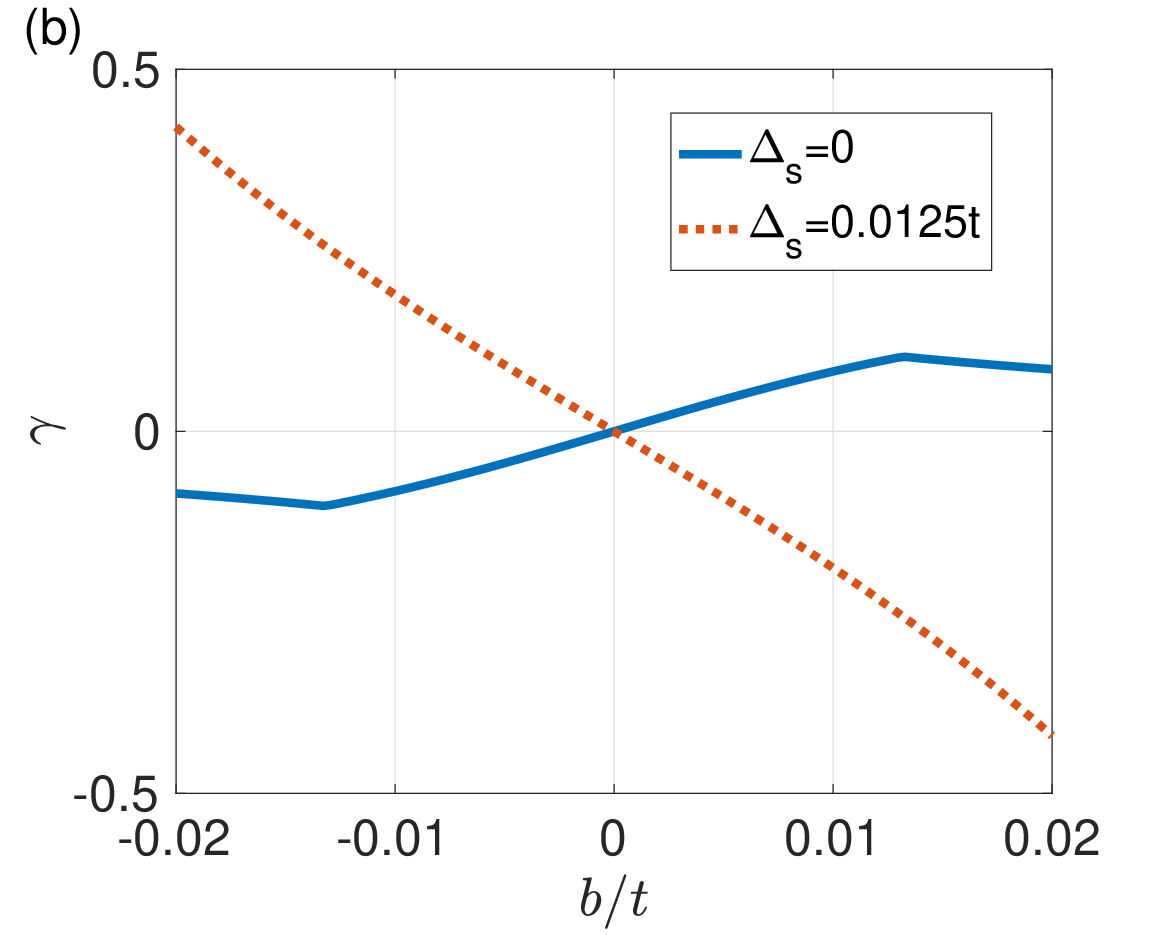}
\caption{(a) Diode effect coefficient $\ga$ versus $\th$ - the angle between the direction of triplet pairing and the direction of SOC. (b) Diode effect coefficient $\ga$ versus $b$ - the Zeeman energy. Legends indicates the value of $\De_s$. In (a), Ordinate for the curve with $\De_s=0.0125t$ ($\De_s=0$) is on the left (right). Other parameters: $\De_t=0.1t$, $\mu_s=\mu_0=-1.875t$, $\al=0.05t$, $b=0.015t$, $L_s=L_q=20$. }\label{fig:ga-th}
\end{figure}

We then explore the dependence of the diode effect coefficient $\gamma$ on the angle ($\theta$) between the direction of triplet pairing and the direction of the SOC in the central quantum wire. In Fig.~\ref{fig:ga-th}(a), we plot $\gamma$ versus $\theta$ for two cases: $\Delta_s = 0$ and $\Delta_s = 0.0125t$, while keeping $\Delta_t = 0.1t$, $\mu_s = \mu_0 = -1.875t$, $\alpha = 0.05t$, $b = 0.015t$, and $L_s = L_q = 20$.
While the diode effect is absent for $\theta = \pi/2$ in the case where $\Delta_s = 0$, this is not true when $\Delta_s = 0.0125t$. For $\theta = \pi/2$, the triplet pairing takes the form $(|\uparrow\downarrow\rangle + |\downarrow\uparrow\rangle)$, meaning the pairing occurs between electrons and holes of opposite spins. As discussed earlier (see Fig.~\ref{fig:phase-eh}(a)), the dynamical phases acquired by electron-hole pairs of opposite spins are identical for states carrying current in both forward and backward directions, which makes us expect that the diode effect coefficient should vanish. 
However, when a singlet pairing term is also present, the total pairing term consists of contributions from both the singlet and triplet pairings. In $k$-space, this total pairing term takes the form $[\Delta_s(|\uparrow\downarrow\rangle - |\downarrow\uparrow\rangle) + \Delta_t \sin{ka}(|\uparrow\downarrow\rangle + |\downarrow\uparrow\rangle)]$. Consequently, in the $|\uparrow\downarrow\rangle$ sector, the pairing strength is $(\Delta_s + \Delta_t \sin{ka})$, while in the $|\downarrow\uparrow\rangle$ sector, it is $(-\Delta_s + \Delta_t \sin{ka})$. The presence of $\sin{ka}$ in the pairing term leads to  pairing strengths having different magnitudes for modes  in the forward and backward directions, breaking inversion and thereby resulting in a diode effect in presence of a time reversal breaking Zeeman field. 
An implication of this argument is that when $\theta \neq 0, \pi$, mixed pairing in the superconductors can lead to a diode effect even in  absence of SOC ($\alpha = 0$) in the central quantum wire, though a nonzero Zeeman field is still required. Our calculations confirm that this indeed holds true. In the last situation, time reversal is broken, but inversion is not broken in the central quantum wire. But both the symmetries are broken in the superconductors.  This means, breaking of time reversal in the central quantum wire is sufficent to produce diode effect if time reversal and inversion are broken in the two superconductors. 
Next, we study the dependence of $\gamma$ on the Zeeman energy $b$ for two cases $\De_s=0, 0.0125t$, fixing other parameters same as earlier. It can be seen from Fig.~\ref{fig:ga-th}(b), that $\gamma$ varies linearly for small $b$. The slopes of the two curves have different signs. This is not surprising since qualitatively, in the purley triplet case, the pairing term mixes electrons and holes of the same spin, while in the mixed pairing, electron of any given spin mixes with holes of both the spins. 

\begin{figure}[htb]
\includegraphics[width=6cm]{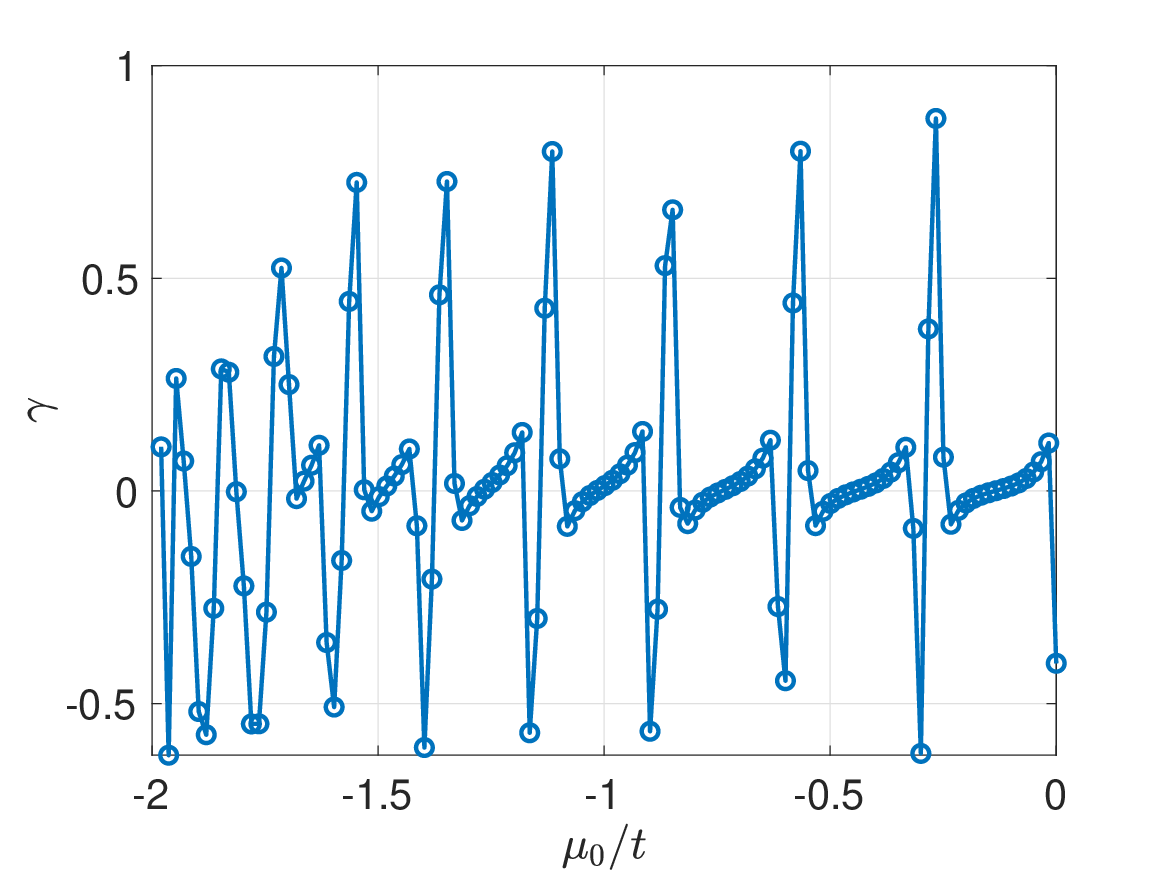}
\caption{Diode effect coefficient $\ga$ versus $\mu_0$ - the chemical potential of the central quantum wire. Oscillations in $\ga$ are reminiscent of Fabry-P\'erot interference.  Parameters: $\mu_s=-1.875t$, $\al=0.05t$, $b=0.015t$, $\th=0$, $L_s=L_q=20$, $\De_s=0.0125t$, $\De_t=10\De_s$.}~\label{fig:fp}
\end{figure}

Now, we investigate the dependence of diode effect coefficient on the chemical potential of the central quantum wire (which can be varied in an experiment by an applied gate voltage).  Previous studies have shown that tuning the chemical potential in a normal metal connected to superconductors leads to oscillations in Josephson current~\cite{sahu23}. This phenomenon arises from Fabry-Pérot interference between plane wave modes within the metallic region~\cite{sahu23,soori12,soori17,soori19,soori22car,das2024car}.

Approximating the dispersion for electrons in the central quantum wire as $E=-2t\cos{ka}-\mu_0$ (since $\al,b\ll t$) (where $a$ is the lattice spacing), we can calculate the corresponding wavevectors $k_{0,j}$ at the peak locations in Fig.~\ref{fig:fp}, taking $E=0$ (since states within the gap contribute dominantly to Josephson current). These wavevectors should satisfy the condition $(k_{0,j+1}-k_{0,j})(L_q+1)a=\pi$. Our analysis of the data in Figure 1 confirms this relationship, with values of $(k_{0,j+1}-k_{0,j})(L_q+1)a/\pi$ approximately equal to 1.09, 0.99, 0.97, 0.99, 1.02, 1.01, and 1.02. The small deviations from 1 are possibly due to the approximations: (i) in the dispersion relation, we neglected $\al$ and $b$, and (ii) while finding $k_{0,j}$, we approximated the energy $E$ to zero. This evidence strongly supports the conclusion that the observed oscillations originate from Fabry-Pérot interference.

{\it Conclusion.-}
We demonstrated that triplet pairing in superconductors can lead to JDE in systems where a spin-orbit-coupled quantum wire, under the influence of a Zeeman field, is connected between two superconductors. Importantly, JDE is absent in such setups when the superconductivity is purely singlet, making this a potential method to probe the presence of triplet superconductivity. Notably, when the directions of the triplet pairing and the Zeeman field in the central quantum wire are non-collinear, SOC in the quantum wire is not necessary to observe JDE, provided that the superconductors host both singlet and triplet pairings.

In the presence of both SOC and a Zeeman field in the quantum wire, JDE is absent when both of the following conditions are satisfied: (i) the pairing is purely triplet, and (ii) the directions of the SOC and the triplet pairing are perpendicular. As a function of Zeeman field, the diode effect coefficient varies linearly  taking the value $0$ at zero Zeeman field.  The chemical potential of the central quantum wire, which can be tuned via an applied gate voltage, causes the diode effect coefficient to oscillate. These oscillations arise from Fabry-P\'erot interference of the plane wave modes within the quantum wire. Therefore, quantum wires can be effectively utilized in Josephson junctions to probe triplet pairings in superconductors through the observation of JDE. Notably, the Josephson diode effect in our study is always accompanied by the anomalous Josephson effect. Josephson junctions incorporating a spin-orbit coupled quantum wire proximitized with a ferromagnetic insulator have been explored experimentally~\cite{raz2022}, making the predictions of our study feasible to test with current technology. Determining current-voltage characteristics and AC Josephson effect of  the setup proposed in this work are some possible topics for future work. 

\acknowledgements 
The author thanks Dhavala Suri, Anindya Das and Bijay Kumar Sahoo for discussions. The author is grateful to Bijay Kumar Sahoo for comments on the manuscript.  The author thanks  SERB Core Research Grant (CRG/2022/004311) and  University of Hyderabad Institute of Eminence PDF for financial support.

\bibliography{ref_trdiode}

\end{document}